\documentclass[aps,prl,preprint,twoside,floatfix,amsmath,showpacs,footinbib]{revtex4-1}

\usepackage{graphicx}
\usepackage{natbib}
\usepackage{color}
\usepackage[normalem]{ulem}

\begin{document}

\newcommand{\pdcoo}{PdCoO$_2$}
\newcommand{\ptcoo}{PtCoO$_2$}
\newcommand{\pdcro}{PdCrO$_2$}
\newcommand{\coo}{CoO$_2$}
\newcommand{\naxcoo}{Na$_x$CoO$_2$}
\newcommand{\nancoo}[1]{Na\ensuremath{_{#1}}CoO$_2$}
\newcommand{\prbacuo}{PrBa$_2$Cu$_4$O$_8$}
\newcommand{\cdi}{CdI$_2$}
\newcommand{\tis}{TiS$_2$}
\newcommand{\mos}{MoS$_2$}
\newcommand{\wse}{WSe$_2$}
\newcommand{\amo}{AMO$_2$}
\newcommand{\twoonefour}{Sr$_2$RuO$_4$}

\newcommand{\mum}[1]{\ensuremath{#1\,\mu\textnormal{m}}}
\newcommand{\sovert}[1]{\ensuremath{#1\,\mu\textnormal{V K}^{-2}}}
\newcommand{\seebeck}[1]{\ensuremath{#1\,\mu\textnormal{V K}^{-1}}}
\newcommand{\resist}[1]{\ensuremath{#1\,\mu\Omega\,\textnormal{cm}}}
\newcommand{\mJmolK}[1]{\ensuremath{#1\,\textnormal{mJ mol}^{-1} \textnormal{K}^{-2}}}
\newcommand{\lorenzunits}{\ensuremath{\,\textnormal{W \Omega K}^{-2}}}
\newcommand{\wkm}[1]{\ensuremath{#1\,\textnormal{W K}^{-1}\textnormal{m}^{-1}}}
\newcommand{\dos}{DOS}
\newcommand{\gfac}{\ensuremath{g_{\textnormal{eff}}}}

\newcommand{\ie}{{\em i.e.}}
\newcommand{\eg}{{\em e.g.}}

\newcommand{\replace}[2]{\sout{#1} \textcolor{red}{#2}}
\newcommand{\tred}[1]{\textcolor{red}{#1}}
\newcommand{\tblue}[1]{\textcolor{blue}{#1}}


\title{Large anisotropic thermal conductivity of intrinsically two-dimensional metallic oxide \pdcoo{}}

\author{Ramzy~Daou}
\affiliation{Laboratoire CRISMAT UMR6508, CNRS / ENSICAEN / UCBN, 6 Boulevard du Mar\'echal Juin, F-14050 Caen, France}
\author{Raymond~Fr\'esard}
\affiliation{Laboratoire CRISMAT UMR6508, CNRS / ENSICAEN / UCBN, 6 Boulevard du Mar\'echal Juin, F-14050 Caen, France}
\author{Sylvie~H\'ebert}
\affiliation{Laboratoire CRISMAT UMR6508, CNRS / ENSICAEN / UCBN, 6 Boulevard du Mar\'echal Juin, F-14050 Caen, France}
\author{Antoine~Maignan}
\affiliation{Laboratoire CRISMAT UMR6508, CNRS / ENSICAEN / UCBN, 6 Boulevard du Mar\'echal Juin, F-14050 Caen, France}

\date{\today}


\begin{abstract}

The highly conductive layered metallic oxide \pdcoo{} is a near-perfect analogue to an alkali metal in two dimensions. It is distinguished from other two-dimensional electron systems where the Fermi surface does not reach the Brillouin zone boundary by a high planar electron density exceeding $10^{15}$\,cm$^{-2}$.
The simple single-band quasi-2D electronic structure results in strongly anisotropic transport properties and limits the effectiveness of electron-phonon scattering.  Measurements on single crystals in the temperature range from 10-300\,K show that the thermal conductivity is much more weakly anisotropic than the electrical resistivity, as a result of significant phonon heat transport. The in-plane thermoelectric power is linear in temperature at 300\,K and displays a purity-dependent peak around 50\,K. Given the extreme simplicity of the band-structure, it is possible to identify this peak with phonon drag driven by normal electron-phonon scattering processes.

\end{abstract}

\maketitle


Artificially created two dimensional electron systems are an ideal testing ground to investigate electronic transport and correlations. The discovery of exotic physical phenomena in the two dimensional electron gas, such as protected edge states and fractionalized quasiparticles in the quantum Hall effects \cite{Klitzing1980,Tsui1982}, motivated the development of ever-cleaner heterostructures. More recently, the discovery of layered, van-der-Waals bonded materials that naturally exhibit two-dimensional electron transport, such as graphene and the dichalcogenides (e.g. \mos{}, \wse{}), has led to the observation of relativistic fermions \cite{Novoselov2005} and topologically protected surface states \cite{Roushan2009}. These systems have in common a low electron density, usually around $10^{12}$\,cm$^{-2}$, and many of the exotic phenomena arise from dominance of the Coulomb interaction over the kinetic energy. As the electronic density is increased towards the metallic state and screening becomes more effective, the kinetic energy becomes more important. However, two-dimensional materials in the limit of high electronic density ($\sim10^{15}$\,cm$^{-2}$, corresponding to one carrier per atomic site) also manifest unusual ground states, as can be seen in the unconventionally superconducting layered materials \twoonefour{} \cite{Maeno1994} or the high-$T_c$ cuprates \cite{Bednorz1986}.

So far missing from the inventory of two-dimensional metals has been the canonical equivalent of the alkali metals: a clean material with a dense electron fluid giving rise to a single cylindrical Fermi contour that avoids the Brillouin zone boundary. Such a material would be topologically suited to fundamental studies of properties that depend strongly on electron-phonon scattering processes, particularly electrical and thermal transport, without the added complication of interband or umklapp scattering.

The delafossite oxide \pdcoo{} is the closest known realisation of this archetype. It is remarkable for its extremely low in-plane resistivity and highly anisotropic character; it is a better conductor than pure palladium metal at room temperature, but only for the in-plane directions \cite{Takatsu2007}. It consists of layers 
of triangularly-coordinated Pd atoms separated by layers of edge-sharing CoO$_6$ octahedra, identical to those that produce strongly correlated behavior in \naxcoo{} \cite{Wang2003} and related misfit cobalt oxides \cite{Limelette2006}. Here, however, the cobalt ions are all in the low spin $3+$ configuration and these layers are effectively inert \cite{Noh2009}.

The Fermi surface measured by quantum oscillations consists of a single warped hexagonal cylinder that never approaches the Brillouin zone boundary and contains precisely one electron per Pd \cite{Hicks2012}, equivalent to a planar carrier density of $1.4\times10^{15}$\,cm$^{-2}$. A striking consequence of this simple Fermi surface contour is that electron-phonon umklapp scattering is activated, producing an exponential temperature dependence of the resistivity $\rho \sim e^{-T_U/T}$ \cite{Hicks2012} with $T_U=160$\,K to at least 30\,K. This behavior has only previously been seen in alkali metals \cite{Bass1990}, but in potassium, for example, $T_U$ is only 29\,K, confining the unusual temperature dependence to $T<4$\,K.
Umklapp processes are the principal way in which resistance to electronic transport is generated at high temperature in clean materials. In their absence phonon drag, whereby the momentum supplied by the external field (electrical or thermal) is conserved by the electron-phonon system, becomes particularly visible in thermal transport. 

In this Letter, we present measurements of the thermal conductivity and thermoelectric power of the model 2D metallic material \pdcoo{} in the temperature range 10-300\,K. We find that the thermal conductivity is strong for a metallic system, reaching values of up to 30 at low temperature.
The thermoelectric power shows a strongly purity-dependent phonon drag peak at low temperature, which allows us to identify the dominant electron-phonon scattering mechanism.

Single crystals of \pdcoo{} were grown by the metathetical reaction as described in Ref.~\onlinecite{Shannon1971}. Platelets of typical dimension $0.5 \times 0.5 \times 0.04$\,mm$^3$ were obtained.
Three $ab$-plane samples were cut from platelets into rectangular bars of approximate dimensions $0.5 \times 0.2 \times 0.04$\,mm.
Electrical contacts were made by bonding \mum{25} gold wires to the samples using Dupont 6838 silver epoxy cured at 180\,$^\circ$C for 5\,min.
Sample dimensions were measured after contacts were applied in a scanning electron microscope. The largest uncertainty arises from the width of the voltage wire contacts, typically 0.05\,mm. 

\begin{figure}
\includegraphics{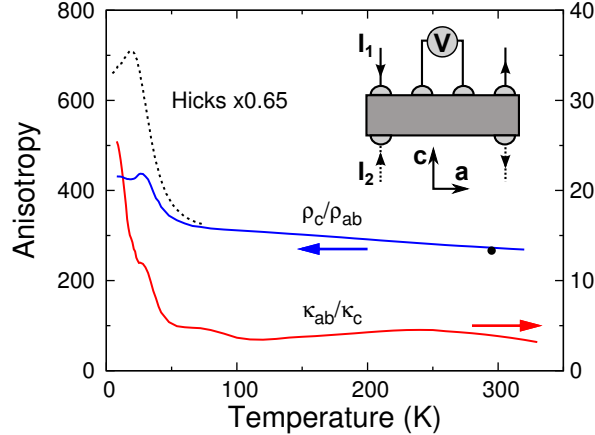}
 \caption{(color online) Anisotropic transport in \pdcoo{}. The blue solid line shows the electrical anisotropy $\rho_c/\rho_{ab}$ (left scale). The black point and dashed black line are the data from Ref.~\onlinecite{Hicks2012} scaled by 0.65. The temperature dependence is similar from 60-300\,K but below this temperature the anisotropy does not increase as much as previously reported. 
 The solid red line shows the thermal anisotropy $\kappa^{tot}_{ab}/\kappa^{tot}_{c}$ (right scale). This is much lower than the electrical anisotropy, as a result of isotropic thermal transport by phonons. The inset shows the contact configuration for sample 1.}
\label{fig1}
\end{figure}

Thermal conductivity and thermopower were measured simultaneously in a custom-built sample holder installed in a PPMS cryostat using a steady-state technique including a calibrated heatpipe to account for thermal losses (principally to radiation) at high temperatures by measuring the input power to the sample \cite{Allen1994}. This loss was negligible below 150\,K. Additional calculated radiation losses were less than 10\% of the measured power at 300\,K. The temperature difference across the samples was measured using chromel/phosphor-bronze thermocouples made from \mum{25} wire directly attached to the samples with Dupont 6838 silver epoxy. The thermoelectric voltage was measured using the phosphor-bronze reference wires, which give an extremely low background contribution ($<$\seebeck{0.05} for $T<80$\,K, as measured against a superconducting reference sample). This value has not been subtracted from the data. Electrical resistivity was measured {\em in situ} by DC current reversal with a typical sensing current of 1\,mA switching at 5\,Hz. 

In order to be sensitive to the strongly anisotropic transport properties of \pdcoo{},  contacts (both electrical and thermal) were made on both ab-surfaces of sample 1 without any overspill of silver epoxy along the exposed ac-surfaces. Current or heat was injected and extracted on either the upper or lower surface and the voltage or temperature differences were always measured on the upper surface (Fig.\ref{fig1} inset). A standard four contacts were made on samples 2 and 3.

A finite element model of sample 1 was used to model anisotropic current flow in the two current injection configurations. In this way the effects of finite contact size and precise placement are taken into account. The measured values are generally a non-linear function of the conductivity anisotropy, as shown by Montgomery \cite{Montgomery1971}. Resistivity measurements provide the ideal test of this arrangement as the anisotropy of the resistivity of \pdcoo{} has already been studied as a function of temperature.

%

The resistivity anisotropy so extracted is $\rho_c/\rho_{ab} = 280$ at room temperature, rising to 450 at low temperature (Fig.~\ref{fig1}). This compares to reported anisotropy values of 400 at room temperature rising to over 1000 at low temperature \cite{Hicks2012}. The high temperature discrepancy can most likely be ascribed to large absolute uncertainty in $c$-axis resistivity measurements arising from unfavorable geometric factors. At low temperatures $\rho_{ab}$ may depend more strongly on sample quality than $\rho_c$, so this difference is also reasonable. The residual in-plane resistivity $\rho_{ab,0}$ extracted from the measurements here is \resist{0.022(3)} compared to \resist{0.009} as reported in Ref.~\onlinecite{Hicks2012}, while at 300\,K $\rho_{ab}=\resist{3.1}$, compared to \resist{2.6}. The temperature dependences of $\rho_{ab}$ and $\rho_c$ are compatible with previous measurements that could be fitted with Einstein contributions to the electron-phonon scattering \cite{Takatsu2007}.

\begin{figure}[t]
\includegraphics{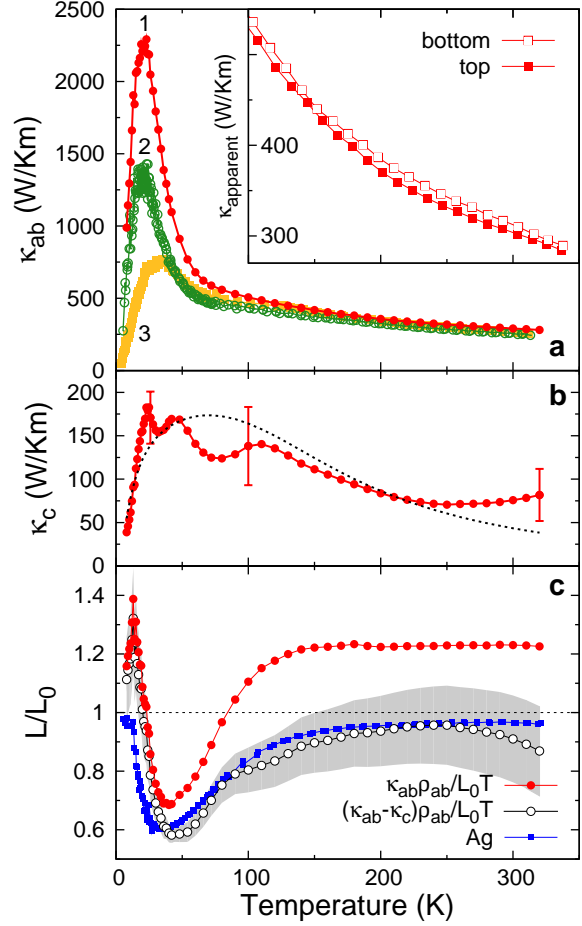}
 \caption{(color online) Anisotropic thermal conductivity of \pdcoo{}. \textit{a)} in-plane thermal conductivity of three samples. The inset shows the raw data from the two contact configurations of sample 1. \textit{b)} out-of-plane thermal conductivity for sample 1 as extracted from the finite element model. The typical uncertainty is indicated by error bars at three points. The dashed black line is the fit to Eq.~\ref{eqn:callaway}. $\kappa^{el}_{c}$ (as estimated from the Wiedemann-Franz law) never exceeds \wkm{6}. \textit{c)} Normalized Lorenz ratio for in-plane transport. The red solid circles are the Lorenz ratio using the raw in-plane data, which is expected to reach unity at high and low temperatures. Subtracting the phonon contribution to the thermal transport (open circles; gray shaded area indicates the uncertainty which arises from the error in $\kappa_c$ at high T and $\rho_{ab}$ at low T) leads to a curve that is much more reminiscent of an ordinary metal at high temperature. For comparison the blue squares show the Lorenz ratio for a silver sample measured using the same apparatus.}
\label{fig2}
\end{figure}

In the case of thermal conductivity measurements, data for the two Montgomery-like configurations are taken consecutively with the resistivity measurements without changing contacts. The directly measured thermal conductances (before the finite element model is employed to extract the thermal conductivity tensor) are very similar in the range 100-300\,K, with only a small systematic difference of between 2-3.5\% (Fig.~\ref{fig2}(a) inset). This difference is close to the resolution of the thermal transport measurement and this results in a reasonably large uncertainty in $\kappa^{tot}_{c}$. The multiple peaks and upturn at high temperature are therefore probably artifacts. The measured excess becomes more pronounced at low temperature, reaching a maximum of around 60\%. 
Due to the strong non-linearity of the Montgomery configuration, this corresponds to a thermal conductivity anisotropy, $\kappa^{tot}_{ab}/\kappa^{tot}_{c} \approx 3-5$, which rises to 30 at low temperature. The anisotropy is shown in Fig.~\ref{fig1} and the thermal conductivities in Fig.~\ref{fig2}.

A rather high value of $\kappa^{tot}_{ab}$ = \wkm{250} at 300\,K (Fig.~\ref{fig2}(a)) is comparable to the noble metals and much higher than both pure Pd metal (\wkm{72}) and the related \naxcoo{} (\wkm{<20}) which shares the same \coo{} layers but has a disordered network of Na$^+$ between them \cite{Foo2004}. 
The thermal conductivity consists of two contributions, arising from heat transported by electronic quasiparticles and phonons, $\kappa^{tot} = \kappa^{el} + \kappa^{ph}$. In the elastic scattering limit reached at low and high temperatures, the Wiedemann-Franz law (WFL) requires that $\kappa^{el} = L_0T/\rho$, where $L_0$ is the Sommerfeld constant. 
Using the WFL, we estimate that $\kappa_{c}^{el}$ is never greater than \wkm{6}. We conclude that $\kappa^{tot}_{c} \approx \kappa^{ph}_{c}$ is dominated by phonon heat transport (Fig.~\ref{fig2}(b)). Thermal transport by phonons is usually rather isotropic. For example, \tis{} has the same \cdi{}-structure layers but they are held together by relatively weak van der Waals forces. We might expect therefore, that heat transport perpendicular to these layers would be strongly inhibited, however, single crystal measurements show that the thermal conductivity anisotropy is 
only a factor of 2 at 300\,K 
\cite{Imai2001}. In \pdcoo{} where the Pd atoms bond the layers together strongly, we would expect the anisotropy in the phonon heat transport to be even weaker. 

This is supported by the dimensionless Lorenz ratio of the thermal to the electrical conductivity, which according to the WFL, should tend to unity as the temperature approaches the Debye temperature $\Theta_D$. This ratio is plotted in Fig.~\ref{fig2}(c) for the in-plane direction. It exceeds unity by $\sim 20$\% at 300K. When $\kappa^{ph}$ is subtracted, however, the ratio drops below unity, as is expected for a normal metal with a reasonably high $\Theta_D = 340$\,K. For comparison the Lorenz ratio of a silver sample (99.99\% pure, $RRR\sim 200$, $\Theta_D = 220$\,K) is also shown; the behavior is very similar. The excess of $L/L_0$ at low temperatures ($<$20\,K) appears to be quite large, however the error here approaches $\pm 15$\%. This should be studied by a more sensitive low-temperature measurement.

We will return to the discussion of the thermal conductivity after examining the in-plane thermoelectric power, $S_{ab}$. It is quasi-linear, small and positive at high temperature, quite different from previous results on polycrystals \cite{Hasegawa2002}. There are typically two contributions to the thermopower in clean metals, arising from electronic diffusion ($S_d \sim T$) and phonon drag ($S_g$, largest at $\Theta_D/5$).

While the sign of $S_{ab}$ is not that expected from a single band electron-like Fermi surface, both the sign and magnitude are in agreement with high temperature transport calculations which assume a constant, energy-independent relaxation time $\tau(\epsilon,k) = \tau_0$ \cite{Ong2010}. A constant $\tau_0$ was also used to successfully model the anisotropy of  magnetoresistance \cite{Takatsu2013}. This suggests that the positive sign is the result of an inverted energy dependence of the band-structure, since:
\begin{equation}
 S = \frac{
   \frac{1}{eT} \int \tau(\epsilon,k) v^2 (\epsilon-\mu)\frac{\partial f}{\partial \epsilon} d\epsilon
  }{
   \int \tau(\epsilon,k) v^2 \frac{\partial f}{\partial \epsilon} d\epsilon
  }
\end{equation}
where $f$ is the Fermi distribution. 

\begin{figure}
\includegraphics{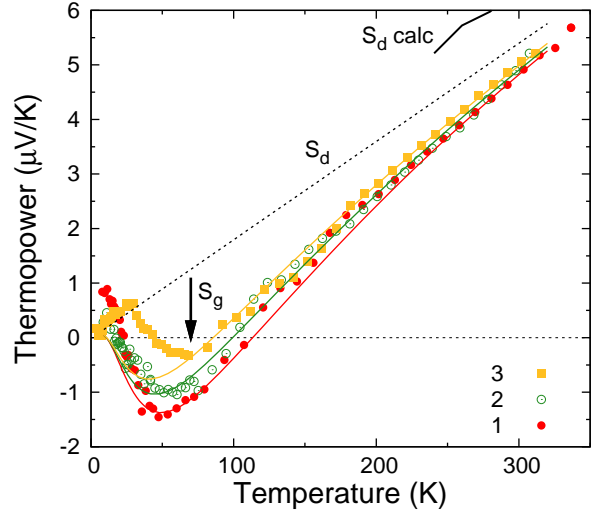}
\caption{(color online) In-plane thermoelectric power of \pdcoo{}. The diffusion thermopower ($S_d$) is linear at high temperatures and the same for all samples. The calculated curve from Ref.~\onlinecite{Ong2010} appears above 240\,K. At low temperature the negative peak is suppressed as the impurity level increases. This is characteristic of a phonon drag ($S_g$) contribution to the thermopower. The lines are fits to Eq.~\ref{eqn:drag}.}
 \label{fig3}
\end{figure}

If $S_d$ changes sign at 100\,K, this must be the result of a change in sign of the sampling function $(\epsilon-\mu)\frac{\partial f}{\partial \epsilon}$. This might be achieved if some part of the band-structure has a strong energy dependence within $k_BT$ of the Fermi energy, as was suggested for the highly warped hexagonal corners \cite{Ong2010}. On the other hand, de Haas-van Alphen results imply that this warping is of a different character than that anticipated by most band-structure calculations \cite{Eyert2008,Ong2010}, arising from direct overlap between Pd 5s states \cite{Hicks2012}. These are not necessarily incompatible observations, given that de Haas-van Alphen is a cold probe of Fermi surface geometry and should not be sensitive to the band-structure 100\,meV from $E_F$. A transport calculation based on the band-structure calculated with a finite $U$ on the Co site \cite{Hicks2012}, compatible with the de Haas-van Alphen results, is currently lacking. This cannot, however, account for the strong sample dependence seen in the thermopower for $T<100$\,K.

Phonon drag is the typical mechanism by which non-monotonic temperature dependence of the thermopower can arise for temperatures around $\Theta_D/5$. It arises when the phonon distribution is out of equilibrium due to a lack of resistive process that dissipate momentum. $S_g$ is visible in most clean metals around $\Theta_D/5$, the canonical example being the noble metals, where a large peak in $S(T)$ is visible around 50\,K \cite{Foiles}.

Ziman pointed out that $S_g$ can be of either sign, depending on whether electron-phonon scattering is dominated by normal or umklapp processes \cite{Ziman1959}. In the former case, a negative $S_g$ (for electrons) arises as phonon momentum acquired from the thermal gradient is transmitted to the electron system via e-p forward scattering, resulting in an additional build-up of negative charge at the cold end of the sample. In the latter case, e-p backscattering is dominant and a positive $S_g$ is the result. It is the former case which applies to \pdcoo{}, where e-p umklapp scattering is strongly suppressed at low temperature.

A key feature of $S_g$ in clean metals is a strong dependence on sample purity. Higher peaks in $\kappa(T)$ imply a greater degree of sample purity (Fig.~\ref{fig2}(a)), and the trend is matched by more strongly negative peaks in $S_g$ (Fig.~\ref{fig3}).
The magnitude of the peak in $S_g$ depends on how far out of equilibrium the electron-phonon distribution can be. Impurity and boundary scattering, as well as phonon-phonon umklapp scattering, transfer momentum to the lattice and so restore thermal equilibrium, thus limiting the proportion of electron-phonon scattering relative to the total scattering rate. Within the Debye approximation, the phonon drag thermopower for free electrons and normal scattering is given by~\cite{Blatt1976}:
\begin{equation}
 S_g =-3\frac{k_B}{e}\biggl(\frac{T}{\Theta_D}\biggr)^3 
      \int_0^{\Theta_D/T} \frac{x^4e^x}{(e^x-1)^2} 
      \frac{\Gamma_{pe}}{\Gamma_{pe} + \Gamma_{po}} dx
\label{eqn:drag}
\end{equation}
where $\Gamma_{pe,po}$ are the scattering rates for phonons by electrons and other processes respectively. As the impurity level increases, the contribution of $\Gamma_{pe}$ becomes relatively smaller, and $S_g$ decreases. 
This formulation is the result of applying the Debye and relaxation-time approximations to the full variational equation derived by Bailyn~\cite{Bailyn1967} and allows us to estimate $S_g$ without detailed knowledge of the phonon dispersion relation, or the momentum dependence of all the types of scattering including the electron-phonon interaction, generally information that can only be derived through complex \textit{ab initio} calculations.

We can access $\Gamma_{po}$ by fitting $\kappa_c^{ph}$ to a Debye-Callaway model as follows~\cite{Callaway1959}:
\begin{equation}
 \kappa^{ph}_c = 3Nk_Bv_{ph}^2 \biggl(\frac{T}{\Theta_D}\biggr)^3
  \int_0^{\Theta_D/T} \tau \frac{x^4e^x}{(e^x-1)^2}  dx
\label{eqn:callaway}
\end{equation}
where $\tau^{-1} = \Gamma_{po} = A_1x^4T^4 + Bx^2T^5e^{-\Theta_D/bT}+v_{ph}/d$ is the scattering rate arising from the phonon-impurity, phonon-phonon and phonon-boundary processes. 
Other formulations, particularly of the phonon-phonon scattering rate, have been explored in the literature. For a good summary, see Ref.~\cite{Tritt2004}. We here fit the data with a minimum of adjustable parameters.
$d=125$\,$\mu$m is the average dimension of the sample, $v_{ph}=3650$\,ms$^{-1}$ is the sound velocity and the constants are found to be $A_1=99$\,s$^{-1}$\,K$^{-4}$, $B=0.067$\,s$^{-1}$\,K$^{-5}$ and $b=4$. The fit is shown in Fig.~\ref{fig2}(b). These parameters are then used to evaluate Eq.~\ref{eqn:drag} where $\Gamma_{pe} = Cx$ is used to describe phonon-electron scattering \footnote{The functional form of $\Gamma_{pe}$ will in general depend sensitively on the details of the Fermi surface and we have chosen the simplest function that provides a reasonable fit over a wide temperature range. Klemens proposed the very similar $\Gamma_{pe} \sim q \sim xT$ for a spherical Fermi surface \cite{Klemens1956}. Scattering of 3D phonons from a 2D Fermi surface will likely have a different parameterization}.
$C = 2.0\times 10^9$ is found to fit the thermopower data for sample 1 in the range 30-300\,K very well when $S_d/T = 18$\,nV\,K$^{-2}$ \footnote{Another estimate of $S_d$ is obtained using the Mott formula for the thermopower expressed in the low temperature limit \cite{Behnia2004}, $S_d/T = \pm\frac{\pi^2k_B^2}{2e\varepsilon_F} = 16$\,nV\,K$^{-2}$ using experimental quantities and assuming parabolic dispersion, which agrees very well with the value obtained.}. Since $A$ is proportional to the impurity concentration, it is varied to fit the other samples, resulting in $A_2 = 129$\,s$^{-1}$\,K$^{-4}$ and $A_3 = 168$\,s$^{-1}$\,K$^{-4}$. This is in accord with the reduced in-plane thermal conductivity seen in samples 2 and 3 relative to 1. The fits are shown in Fig.~\ref{fig3}. A similar effect was seen when disorder was quenched into gold \cite{Huebener1964}.

At low temperatures ($T<25$\,K) it is clear that there is further sample-dependent structure in the thermopower which is difficult to explain by the phonon drag mechanism, which should decrease monotonically below $\Theta_D/5$. One possible explanation is the presence of a single-ion Kondo effect due to very low-level concentrations of magnetic impurities, which can enhance the thermopower at temperatures comparable to the Kondo scale~\cite{Kopp1975}. A very slight upturn in the $\rho_{ab}$ for $T<10$\,K may be a sign of this \cite{Hicks2012}.

The additional constraint required to see signs of phonon drag in the resistivity is that electron-phonon umklapp scattering should also be suppressed. This results in the suppression of the ordinary electron-phonon resistivity and the exponential temperature dependence at $T\ll T_U$. Returning to the thermal conductivity, the resemblance of $L/L_0$ to an ordinary metal therefore suggests that $\kappa_{ab}^{el}$ must also be relatively enhanced. This implies that small-angle scattering, responsible for the dip in $L/L_0$ at $\Theta_D/5$, has been suppressed to the same degree as the large angle scattering channel responsible for electrical resistivity.

We emphasize that it is only the simplicity of the electronic structure near the Fermi energy which has allowed even such a qualitative interpretation of the data. Even in the alkali metals, however, the temperature dependence and sign of the thermopower does not always follow the simple expectations of Boltzmann transport theory; for example the positive thermopower of lithium can only be explained if the relaxation time approximation is false \cite{Xu2014}. \pdcoo{} is a rather unique material that may well be the closest realization of a textbook 2D metal available in single crystal form. It is not unreasonable to propose that quantitative calculations of electron-phonon scattering could accurately model the drag resistivity and thermopower in this case. \pdcoo{} is additionally remarkable for its elevated thermal conductivity compared to other oxides.

\bibliographystyle{unsrt}

\end{document}